\documentclass[doublecol, figures]{epl2} 

\usepackage{amsmath}
\usepackage{amsfonts}
\usepackage{amssymb}
\usepackage{graphicx}
\usepackage{epstopdf}

\newcommand{\be}{\begin{equation}}
\newcommand{\en}{\end{equation}}

\newcommand{\pre}{Phys. Rev. E }

\newcommand{\avg}[1]{\left< #1 \right>}

\newcommand{\erf}{{\rm erf}}
\newcommand{\at}{\avg{t}}
\newcommand{\as}{\avg{s}}

\newcommand{\te}{E-\avg{t}}

\newcommand{\hrho}{\hat{\rho}}
\newcommand{\hi}{\hat{I}_q}

\title{Eigenvectors under a generic perturbation: non-perturbative results from the random matrix approach}
\shorttitle{Eigenvectors under a generic perturbation} 

\author{K. Truong  \and A. Ossipov}

\institute{                    
   School of Mathematical Sciences, University of Nottingham, Nottingham NG7 2RD, United Kingdom\\
}

\abstract{
We consider eigenvectors of the Hamiltonian $H_0$ perturbed by a generic perturbation $V$ modelled by a random matrix from the
Gaussian Unitary Ensemble (GUE). Using the supersymmetry approach we derive analytical results for the statistics of the eigenvectors, which
are non-perturbative in $V$ and valid for an arbitrary deterministic  $H_0$. Further we generalise them to the case of a random $H_0$,
focusing, in particular, on the Rosenzweig-Porter model. Our analytical predictions are confirmed by numerical simulations.
}

\begin{document}

\maketitle

\section{Introduction}

One of the basic problems in quantum mechanics is to find the eigenvectors of the Hamiltonian $H_0$ perturbed by another Hamiltonian  $V$. The  standard approach to this problem is given by the perturbation theory, which allows us to construct the eigenvectors of 
\be\label{H-def}
H=H_0+V 
\en
as a power series in the matrix elements of $V$. Normally the results of the perturbation theory are valid only if these matrix elements are sufficiently small compared to a typical  level spacing of $H_0$.  In many applications, this restriction is not fulfilled and one needs to look for  non-perturbative solutions, which are usually available only for exactly solvable models. 

In this work we consider a generic perturbation $V$ modelled by a $N\times N$ random matrix from the Gaussian Unitary Ensemble  (GUE) with the variance $\avg{|V_{ij}|^{2}}=1/N$ \cite{Mehta}.  The Random Matrix Theory (RMT) has proven to provide a correct quantitative description of the universal features of complex quantum systems, including  important examples such as chaotic and disordered systems \cite{RMT_book}. Since we are interested in the eigenvectors of $H$ relative to the eigenbasis of the unperturbed Hamiltonian, we assume that the $N \times N$ matrix $H_0$ is diagonal with $(H_0)_{ii}=d_i$. In the first part of the paper we consider $d_i$ as non-random, deterministic energy levels, whose typical level spacing is of order  $1/N$\footnote{Our results below show that this scaling  makes the two matrices $H_0$ and $V$ to be of the same order in $N$.}. Applying the  supersymmetry technique \cite{Efetov} we derive non-perturbative analytical results for the local moments of the eigenvectors  $\avg{|\psi_n|^{2q}}$, which hold true for arbitrary $d_i$ in the limit $N\to \infty$.  The Hamiltonians defined by Eq.\eqref{H-def} with $V$ represented by a random matrix and a non-random $H_0$ are known as random matrix models with external source and they have various applications in network models, telecommunication, neuroscience and other areas.  Their spectral properties have been studied intensively for many years \cite{P72,BHZ95,BH96, BH98,Kh96,BK04, G06, F13, AFM15,GG16}, however we are not aware of any analytical results for their eigenvectors. 

The interest to the properties of eigenvectors in the random matrix models has been renewed recently due to the new questions, which  arose in the context of the  Anderson and many-body localisation \cite{Abrahams, BAA06}.   In particular, a lot of attention has been given to the existence of the non-ergodic delocalised states representing an intermediate phase between the localised and the extended phases of the Anderson transition \cite{BRT12,LAKS14,Kravtsov,G15,PIA16,M16,TMS16,ACIK16,TM16}. Although one cannot expect that toy random matrix models, such as one considered in this work, can capture all the features of the many body localisation, they can provide a detailed quantitative description of some of the features, which might be not accessible for more realistic models. In this context it is natural to consider $H_0$ to be random, so that $d_i$ play the role of the on-site disorder, while $V$ has the meaning of structural disorder. These two types of disorder are common to other models  such as random regular graphs and believed to be essential for the many body localization \cite{Kravtsov}. In the second part of the paper, we assume that $d_i$ are uncorrelated  Gaussian distributed random variables, characterised by $\avg{d_i}=0$ and $\avg{d_i^2}=\sigma^2$ . Using our results derived in the first part as a starting point, we are able to average over $d_i$  and thus find closed analytical formulas for the statistics of the eigenvectors.  In particular,   they can be applied to the Rosenzweig-Porter model \cite{RP60}, for which $\sigma$ scales with $N$ in a non-trivial way $\sigma^2=N^{\gamma-1}$. We find that for $1<\gamma<2$, the eigenvectors are non-ergodic and characterised by non-trivial fractal dimensions in agreement with the recent results from Ref \cite{Kravtsov}.


\section{The model with non-random $H_0$} In order to calculate the moments of the eigenvectors of $H=H_0+V$ we employ the supersymmetry technique \cite{Efetov}. We have recently used the same method for finding the statistics of the eigenvectors of the random matrix $W\tilde{H} W$, where $\tilde{H}$ is a GUE matrix and $W$ is a diagonal matrix with non-random diagonal elements \cite{TO16}.  It turns out that for the present problem one can follow exactly the same steps of the calculations and the only difference is in the dependence of the supersymmetric action on the diagonal matrix elements of $H_0$. For this reason, we refer the reader to Ref.\cite{TO16} for the details of the calculations and the notation is used. The matrix elements of Green's functions at the energy $E\pm \mathrm{i}\epsilon $ can be written as an integral over the supermatrix $Q$ with the action
\be 
S[Q]=\frac{N}{2} \mathrm{Str}\:Q^2 + \sum_{i=1}^N \mathrm{Str}\ln \left[E-d_i-Q+\mathrm{i}\epsilon \Lambda\right],
\en
where $\Lambda = \mathrm{diag}(1,1,-1,-1)$.
In the limit $N\to \infty$ the integral is dominated  by the saddle-points which satisfy the saddle-point equation 
\begin{equation}
Q = \frac{1}{N}\sum_{i=1}^N \frac{1}{E-d_i-Q},
\end{equation}  
whose solutions can be parametrised as follows
\begin{equation}
Q_{s.p.} = t\bm{1}-\mathrm{i}sT^{-1}\Lambda T,\label{saddlepointeqn}
\end{equation}
where  the matrix $T^{-1}\Lambda T$ parametrises the saddle-point manifold  in the absence of $H_0$ \cite{Efetov} and $s\neq 0$ and $t$ are two real parameters satisfying the simultaneous equations
\begin{align}\label{ts-system}
t = \frac{1}{N} \sum_i^N \frac{E-t-d_i}{(E-t-d_i)^2+s^2},\nonumber\\
1 =\frac{1}{N}\sum_i^N \frac{1}{(E-t-d_i)^2+s^2}.
\end{align}
Using the above results the integral over $Q$ can be calculated and various physical quantities, which can be expressed through the Green's functions can be found. In particular, the density of states turns out to be directly related to the parameter $s$
\begin{equation}\label{dos}
\rho(E) = \frac{s}{\pi}.
\end{equation}
The local moments of the eigenvectors  $I_q(n)=\avg{|\psi_n|^{2q}}$, where $\psi_n$ refers to the $n$th component of the eigenvector $\psi$ and $q$ is a positive integer, are given by
\begin{equation}\label{mom-result}
I_q(n) = \frac{1}{N^q}\left[\frac{1}{(E-t-d_n)^2+s^2}\right]^q \Gamma (q+1),
\end{equation}
where $\Gamma (z)$ is the gamma function. It is easy to check that by setting all $d_i=0$ we recover the GUE results. Indeed, in this case  the system \eqref{ts-system} can be easily solved giving $s = \sqrt{1-(E/2)^2}$ and $t = E/2$. Substituting this solution into Eq.\eqref{dos} and Eq.\eqref{mom-result} yields 
\be
\rho^{GUE}(E)= \frac{1}{\pi} \sqrt{1-(E/2)^2},\ I_q^{GUE}(n) =\frac{ \Gamma (q+1)}{N^q},
\en
which are the well-known results for the GUE matrices \cite{Mehta}.

For $H_0\neq 0$, Eq.\eqref{mom-result} implies that $I_q(n)$ depends explicitly only on the corresponding matrix element $d_n$ in a very simple way. At the same time,  there is also an implicit dependence on all $d_i$ through the global parameters $s$ and $t$ demonstrating that the result is non-perturbative.   

The fact that $I_q(n)\propto N^{-q}$ have the same scaling with $N$ as the GUE result means that the eigenvectors, which are completely localised in the absence of the perturbation, become fully extended for any values of $d_i$, i.e. for an arbitrary strength of the perturbation. 
Another manifestation of the same phenomenon is the fact that the perturbation mixes all the levels of $H_0$, as it follows form the expression for $\rho (E)$. At the same time, the ratio $I_q/I_q^{GUE}$ can be arbitrary large implying that the eigenvectors of $H$ can be very different from the GUE eigenvectors. We return to this point in the next section.

Another way to interpret the result \eqref{mom-result} is to consider $H_0$ as a perturbation of $V$. Then the scaling with $N$ of $I_q(n)$ demonstrates that the eigenvectors remain extended at arbitrary perturbation, although the density of states can be drastically changed. Such robustness of the eigenvectors is similar to the universality of the two-point spectral correlation function studied previously for this model \cite{BHZ95,BH96, BK04}.    

\begin{figure}
\onefigure[width=\linewidth]{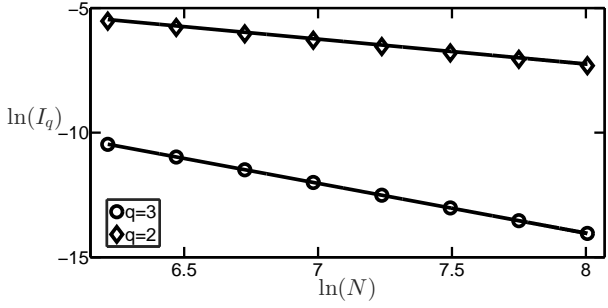}
\caption{The moments of the eigenvectors $I_q=\sum_{n=1}^N I_q(n)$ for $N$ ranging from $500$ to $3000$, where the symbols represent the numerical data and the straight lines represent the analytical results for $d_i = -1+\frac{2}{N}(i-1)$.}
\label{fig.1}
\end{figure}

We test our analytical result \eqref{mom-result} by the numerical simulations for the specific choice of $d_i=-1+\frac{2}{N}(i-1)$.
Using the direct diagonalisation the moments were calculated for the eigenvectors, whose eigenvalues are close to $E=0$. Fig.~\ref{fig.1} shows that the numerical results for $q=2$ and $q=3$ are in excellent agreement with the analytical predictions. 

\section{The model with random $H_0$}
Now we consider the case, where $H_0$ is random: we assume that $d_i$ are independent Gaussian variables with $\avg{d_i}=0$ and $\avg{d_i^2}=\sigma^2$. The parameter $\sigma$ controls the strength of the perturbation: the perturbation is weak (strong) at $\sigma \gg 1$ ($\sigma \ll 1$).
It turns out that the result of the previous section for non-random $H_0$ can be used as a convenient starting point for calculations in this case. Indeed, if $d_i$ are random, then $t$ and $s$  become also random variables, which still can be determined by solving the system of the equations \eqref{ts-system} for each realisation of $d_i$. The structure of Eqs.\eqref{ts-system} suggests that $t$ and $s$ are self-averaging quantities, i.e. their deviations from their mean values vanish as $N\to \infty$. This assumption is confirmed by our numerical simulations. Therefore when we average \eqref{ts-system} over the Gaussian distribution of $d_i$ we can replace $t$ and $s$ by their mean values:
\begin{align}
\at=\avg{\frac{\te-d}{(\te-d)^2+\as^2}}_d,\label{s_t-system-avg-1}\\
1=\avg{\frac{1}{(\te-d)^2+\as^2}}_d,\label{s_t-system-avg-2}
\end{align}
where we write $d$ instead of $d_i$, as all $d_i$ have the same probability distribution function. Using the Fourier transform of the Gaussian distribution function of $d$, we perform the averaging explicitly and obtain the following result

\begin{align}\label{as-eqn}
\at=-\mathrm{i}\sqrt{\frac{\pi}{8}}\frac{1}{\sigma}e^{-\frac{(E-\at+\mathrm{i}\as)^2}{2\sigma^2}}
F_{-}\left(\frac{E-\at}{\sqrt{2}\sigma},\frac{\as}{\sqrt{2}\sigma}\right),\nonumber\\
1=\sqrt{\frac{\pi}{8}}\frac{1}{\sigma \as}e^{-\frac{(E-\at+\mathrm{i}\as)^2}{2\sigma^2}}
F_{+}\left(\frac{E-\at}{\sqrt{2}\sigma},\frac{\as}{\sqrt{2}\sigma}\right),\nonumber\\
F_{\pm}(x,y)=1\pm e^{4\mathrm{i}xy}(1-\erf (\mathrm{i}x+y))+\erf(\mathrm{i}x-y),
\end{align}
where $\erf(z)$ is the error function. Solving numerically this system of  two equations one can find $\as$ and $\at$ and hence the density of states for any values of $E$ and $\sigma$:
\be\label{hrho}
\hrho(E)=\frac{\as}{\pi},
\en
where the different notation for the density of states is used in order to distinguish between the results of this section and the previous one.

In order to derive the expression for the moments we need to average Eq.\eqref{mom-result} over $d_n$ and replace $s$ and $t$ by their mean values. This can be done by differentiating $q-1$ times Eq.\eqref{s_t-system-avg-2} and using the second equation of  Eqs.\eqref{as-eqn}:
\begin{align}\label{hi}
\hi\equiv\sum_n\hi(n)=\frac{q}{N^{q-1}}\left[\left(-\frac{1}{2y}\frac{d}{d y}\right)^{q-1}G(y)\right]_{y=\as},\nonumber\\
G(y)=\sqrt{\frac{\pi}{8}}\frac{1}{\sigma y}e^{-\frac{(E-\at+\mathrm{i}y)^2}{2\sigma^2}}
F_{+}\left(\frac{E-\at}{\sqrt{2}\sigma},\frac{y}{\sqrt{2}\sigma}\right).
\end{align} 
The derivatives can be evaluated explicitly for any positive integer $q$, however the resulting expressions become quite cumbersome. The simplest one for $q=2$ reads
\begin{align}\label{itwo}
\hat{I}_2=\frac{1}{N\as^2}\left[1+\frac{1}{\sigma^2}+H(z)\right]_{z=\frac{E-\at-\mathrm{i}\as}{\sqrt{2}\sigma}},\nonumber\\
H(z)=\mathrm{i}\sqrt{2}\frac{\as}{\sigma}z^{\ast}-\mathrm{i}\sqrt{\frac{\pi}{2}}\frac{E-\at}{\sigma^3}e^{-z^2}\left(1-\erf (\mathrm{i} z)\right).
\end{align} 

Let us analyse the above results in the two opposite limits $\sigma\to 0$ and $\sigma\to \infty$. In the limit $\sigma\to 0$, Eqs.\eqref{as-eqn} can be simplified using the asymptotic expansion of $\erf (z)$ at $z\to \infty$:
\begin{eqnarray}
\at&\approx& \frac{E-\at}{(E-\at)^2+\as^2},\nonumber\\
1&\approx &\left(E-\at\right)^2+\as^2.
\end{eqnarray}
Solving them we obtain
\be
\at\approx \frac{E}{2},\quad \as\approx \sqrt{1-\frac{E^2}{4}},\quad \hrho (E)\approx \frac{1}{\pi}\sqrt{1-\frac{E^2}{4}}.
\en
Thus the density of states is given by the Wigner semicircle, which is expected in the limit $H_0\to 0$.
 
Now consider the opposite case $\sigma\to\infty$. Using the fact $\as/\sigma$ is a small parameter we can simplify Eqs.\eqref{as-eqn} and find the approximate solution:
\begin{align} 
\at \approx \sqrt{\frac{\pi}{2}}\frac{1}{\sigma \mathrm{i}}e^{-\frac{E^2}{2\sigma^2}}\erf\left(\frac{\mathrm{i}E}{\sqrt{2}\sigma}\right),\nonumber\\
\hrho(E)=\frac{\as}{\pi}\approx \frac{1}{\sqrt{2\pi}\sigma }e^{-\frac{E^2}{2\sigma^2}}.
\end{align}
The density of states is Gaussian and determined solely by $H_0$, which is natural to expect in the limit $\sigma\to \infty$.

\begin{figure}
\onefigure[width=\linewidth]{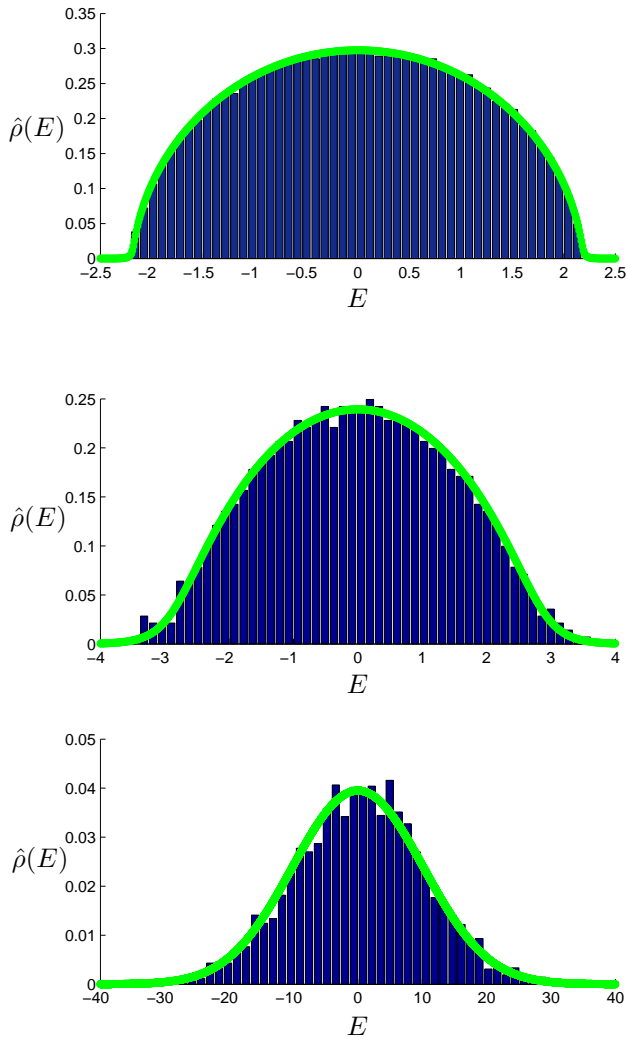}
\caption{The histograms for the density of states, for $\sigma = 0.4,1,10$, calculated for $N=3000$ over 3000 realisations, are compared with  the analytical predictions, calculated from Eqs.(\ref{as-eqn}-\ref{hrho}), represented by the solid lines.}
\label{fig.2}
\end{figure}

For general $\sigma$ the density of states interpolates between these two limiting cases. In Fig.~\ref{fig.2} we can see the transition from the Wigner's semi-circle law to the Gaussian distribution as we increase the value of $\sigma$.  The numerical simulations confirm that the expression for $\hrho(E)$ given by Eqs.(\ref{as-eqn}-\ref{hrho}) is valid for any value of $\sigma$.

Our formula for $I_q$ can be also examined in the two limiting cases. For simplicity, we focus on $I_2$ given by Eq.\eqref{itwo}. In the limit $\sigma\to 0$, using the asymptotic expansion of $\erf (z)$ at $z\to \infty$, we recover the GUE result $\hat{I}_2\approx \hat{I}_2^{GUE}=  1/N$, as expected. This result can be generalised to all values of $q$: $\hi\approx \hat{I}_q^{GUE}$, as $\sigma \to 0$. 

In the opposite limit $\sigma\to\infty$, we obtain 
\be
\hat{I}_2\approx \frac{ 2\sigma^2}{\pi N}, 
\en
which is a less trivial outcome. Indeed, based on the result for the density of states, one could expect that $V$ becomes irrelevant in the limit $\sigma\to \infty$. This would imply that the moments are determined by the completely localised eigenvectors of the diagonal matrix $H_0$ and hence they are $N$-independent.  It is clear that this scenario is correct, provided that  the limit $\sigma\to \infty$ is taken before the limit $N\to \infty$. The $1/N$-dependence of  $\hat{I}_2$ means that the perturbation completely changes the nature of the eigenvectors, if the order of the limits is opposite.

The second moment $\hat{I}_2$, which is known as an inverse participation ratio, gives the inverse of the number of essentially non-zero components of the eigenvectors.  The $1/N$-scaling of  $\hat{I}_2$ shows that the number of such components is of order $N$ and the eigenvectors are ergodic. At the same time $\hat{I}_2/\hat{I}_2^{GUE}\approx2\sigma^2/\pi\gg 1$ meaning that the eigenvectors are less extended than the GUE eigenvectors and they may occupy an arbitrary small, but finite fraction of the whole system.  

\begin{figure}
\onefigure[width=\linewidth]{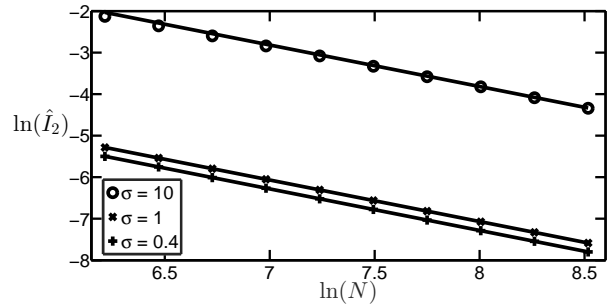}
\caption{ $\hat{I}_2$ calculated for the three different values of $\sigma = 0.4,1,10$ and for $N=500$ to $N=3000$. The symbols represent the numerical results and the solid lines represent Eq\eqref{itwo}.}
\label{fig.3}
\end{figure}

The formula for $\hat{I}_2$ \eqref{itwo} is confirmed by the numerical simulations presented in  Fig.~\ref{fig.3}. The moment $\hat{I}_2$ was calculated  for the eigenvectors, whose eigenvalues are close to $E=0$,  for $N$ ranging from $500$ to $3000$ over $500$ to $1000$
realisations.

\section{The Rosenzweig-Porter model}
The results of the previous section show that the eigenvectors of $H$ are always extended even at arbitrary large, but fixed $\sigma$. In order to make the appearance of non-extended states possible,  one can allow for $\sigma$ to be $N$-dependent. More specifically we consider the Rosenzweig-Porter model, in which  $\sigma^2=N^{\gamma-1}$. The spectral properties of this model were studied intensively in the past \cite{P95,G96,GMG97,KSh98} and it was found that the two-point spectral correlation function undergoes a transition from the Wigner-Dyson to the Poisson form at $\gamma=2$. Motivated by new developments in the many-body localisation,  the statistical properties of the eigenvectors have been investigated  in a recent paper by Kravtsov et al.\cite{Kravtsov}. One of the results of their work was the existence of a new phase transition at $\gamma=1$ separating the ergodic ($\gamma <1$) and non-ergodic ($1<\gamma<2$) states. The ergodic states are similar to the GUE extended states and their moments have the $N^{1-q}$ scaling with the system size. In contrast,  the moments of the non-ergodic states are characterised by non-trivial fractal dimensions $D_q$: 
\be
\hi\propto N^{-D_q(q-1)},
\en
so that they have properties of the multi-fractal critical states typical for the Anderson transition \cite{EM08}. At the same time,  the two-point correlation function of the corresponding eigenvalues  are given by the Wigner-Dyson result, which is a signature of the delocalised eigenvectors.
 
It turns out that the general formula for the moments of the eigenvectors derived in the previous section can be directly applied to the eigenvectors of  the Rosenzweig-Porter model for $\gamma<2$. If $\gamma<1$, then $\sigma\to 0$ as $N\to \infty$, and then  $\hi\approx \hat{I}_q^{GUE}$, as it was discussed above.

For $1<\gamma<2$,  substituting $\sigma=N^{\frac{\gamma-1}{2}}$ into Eq.\eqref{itwo} and keeping only the leading in $N$ term we obtain
\be
\hat{I}_2\approx \frac{ 2}{\pi }N^{\gamma-2}, 
\en
which implies that $D_2=2-\gamma$ in agreement with the result of Ref. \cite{Kravtsov}, which was derived using the perturbation theory supplemented by some heuristic arguments. Our result gives the prediction not only for the exponent of the power law, but also for the constant factor. 

We can notice that the leading contributions to $\hi$  come from the derivatives of the $1/y$ term in \eqref{hi}, therefore we can easily generalise the above result to other values of $q$:
\be\label{Iq-RP}
\hi \approx \frac{ q(2q-3)!!}{\pi^{q-1} }N^{(\gamma-2)(q-1)}, 
\en
confirming again the formula for $D_q=2-\gamma$ for $q>1/2$ from Ref. \cite{Kravtsov}.

For $\gamma>2$, Eq.\eqref{Iq-RP} gives divergent moments signalling that our approach breaks down. In this case, $V$ becomes a small
perturbation compared to $H_0$ and hence the moments can be computed perturbatively using a different approach \cite{FOR09, Kravtsov}.  

\begin{figure}
\onefigure[width=\linewidth]{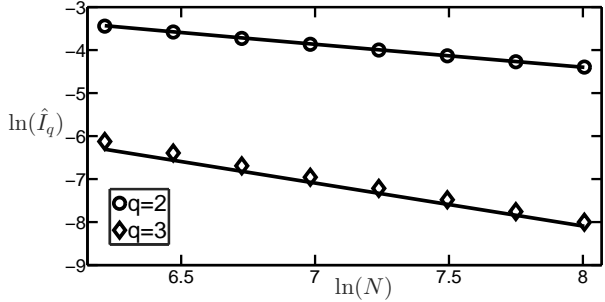}
\caption{Numerical simulation for the Rosenzweig-Porter model for $\gamma=1.5$  and $N$ ranging from $500$ to $3000$. The symbols represent the numerical results and the solid lines represent the analytical prediction.}
\label{fig.4}
\end{figure}

In Fig.~\ref{fig.4} we compare the results of the numerical simulations with  Eq.\eqref{hi} for $\sigma=N^{\frac{\gamma-1}{2}}$ and  $q=2,3$. We used the scaled Hamiltonian $H/\sigma$ in the numerical calculations, because it has the same eigenvectors as $H$, but its eigenvalues remain bounded in the limit $N\to \infty$. The agreement between the numerical results and the analytical predictions is very good in both cases.

\section{Conclusions}
We studied the eigenvectors of the diagonal $N\times N$ matrix   $H_0$ perturbed by a GUE matrix $V$. We consider three different cases: (i) $H_0$ is deterministic, (ii) $(H_0)_{ii}$ are Gaussian distributed with a constant variance $\sigma^2$, (iii) $(H_0)_{ii}$ are Gaussian distributed with the variance  $\sigma^2=N^{\gamma-1}$. Employing the supersymmetry method we derived non-perturbative analytical results for the density of states and the moments of the eigenvectors, which are valid in the limit $N\to\infty$. 

In the first two cases, we found that the initially localised eigenvectors become delocalised at arbitrary weak perturbation. At the same
time, the degree of their ergodicity can be parametrically smaller compared to the completely delocalised  GUE eigenvectors. In the third
case, the eigenvectors can be completely extended, localised or critical depending on the parameter $\gamma$. In particular, for
$1<\gamma<2$, we found that the eigenvectors are non-ergodic and charaterised by non-trivial fractal dimensions, in agreement with the
recent results.  After the completion of this work, we became aware of Ref.\cite{FVB16} and Ref.\cite{M16arxiv}, in which the Green's functions and the eigenvectors of the Rosenzweig-Porter model are investigated by different methods.

\acknowledgments
KT acknowledges support from the Engineering and Physical Sciences Research Council [grant number EP/M5065881/1].

\end{document}